\documentclass[doublecol,linenumbers]{epl2}
\usepackage{graphicx}
\usepackage{subfigure}
\usepackage{amsthm}
\usepackage{amsmath}
\usepackage{amssymb}
\usepackage{verbatim}
\usepackage{dcolumn}% Align table columns on decimal point
\usepackage{bm}% bold math
\usepackage{epsf}
\usepackage{color}
\usepackage[colorlinks=true,citecolor=blue,linkcolor=blue,urlcolor=blue]{hyperref}%
\usepackage{xcolor}
\usepackage{dsfont}
\usepackage{caption}
\usepackage{subcaption}

\usepackage[version=4]{mhchem}

\newcommand{\bla}{bla\\bla\\bla\\bla\\bla}

\newcommand{\mc}[1]{\mathcal{#1}}

\DeclareMathOperator{\Tr}{Tr} 

\newcommand{\te}{\text}

\title{Correcting noisy quantum gates with shortcuts to adiabaticity}

\author{Moallison F. Cavalcante\inst{1,2,3}, Bari\c{s} \c{C}akmak\inst{4}, Marcus V. S. Bonan\c{c}a\inst{2}, \and Sebastian Deffner\inst{1,3,5} \thanks{E-mail: \email{moallison@umbc.edu}}}
\shortauthor{Moallison F. Cavalcante et al.}

\institute{\inst{1} Department of Physics, University of Maryland, Baltimore County, Baltimore, MD 21250, USA\\
\inst{2} Gleb Wataghin Physics Institute, The University of Campinas, 13083-859, Campinas, S\~{a}o Paulo, Brazil\\
\inst{3} Quantum Science Institute, University of Maryland, Baltimore County, Baltimore, MD 21250, USA\\
\inst{4} Department of Physics, Farmingdale State College -- SUNY, Farmingdale, NY 11735, USA\\
\inst{5} National Quantum Laboratory, College Park, MD 20740, USA
}

\abstract{Unitary quantum gates constitute the building blocks of Quantum Computing in the circuit paradigm. In this work, we engineer a locally driven two-qubit Hamiltonian whose instantaneous ground-state dynamics generates the controlled-NOT (CNOT) quantum gate. In practice, quantum gates have to be implemented in finite-time, hence non-adiabatic and external noise effects debilitate gate fidelities. Here, we show that counterdiabatic control can restore gate performance with near perfect fidelities even in open quantum systems subject to decoherence.}

\begin{document}

\maketitle

\section{Introduction}

To achieve genuine quantum advantage, the development of fault-tolerant quantum computers is required \cite{nielsen2000quantum}. This poses significant technological challenges, since in practice any quantum computer has to be built from physical platforms, where qubits are actual physical entities that can be manipulated with the goal of performing unitary gate operations \cite{Sanders2017,BruzewiczAPR2019,Slussarenko2019APR,Wintersperger2023,Ezratty2023EPJA}. Therefore, it is clear that such operations cannot be dissociated from the underlying physical dynamics, which are inherently noisy \cite{Tripathi2025CR}. Significant work has been published on how to correct for such gate errors, employing, for instance, sophisticated optimal control theory \cite{PhysRevA.65.042301,PhysRevA.86.012317,Goerz_2014,Deffner_2014,Du_2016,Santos_2017,Utkan2022PRR,Kanaar2024QST,PRL_URC_2024,Li_2024,Vetter2024,singh2025}, geometric and dynamical principles \cite{Acconcia2015PRE,Walelign2024PRA,Amer2025}, or decompositions into ``trotterized'' or ``digitized'' gates \cite{Hegade_PRA_2021,Hegade2022PRR,Zhao2024PRL}. In this work, we will take an alternative approach that fully exploits the prowess of counterdiabatic driving.

A set of quantum gates that enables the implementation of any arbitrary $N$-qubit unitary operation is defined as a universal quantum gate set~\cite{nielsen2000quantum}. Naturally, such sets are aimed to be composed of only a small number of gates and low Hilbert space dimensions, to reduce complexity and to ease the controllability of physical systems.
%Universal quantum computation is a concept based on the fact that an arbitrary unitary gate operation $U$ can be decomposed into a few unitary gates 
A standard example of a universal gate set for quantum computing is given by the Hadamard, phase, $\pi/8$, and controlled-NOT (CNOT) gates \cite{PRA_UQC_1995,nielsen2000quantum}. The CNOT is a two-qubit gate that operates on a target qubit conditioned on the state of the control qubit. In particular, given the regular two-qubit computational basis states $\{|00\rangle,|01\rangle,|10\rangle,|11\rangle\}$, %the effect of the CNOT gate is as follows
%\begin{align}
%|00\rangle &\rightarrow|00\rangle, & |01\rangle  &\rightarrow|01\rangle, & |10\rangle &\rightarrow|11\rangle, & |11\rangle &\rightarrow|10\rangle.
%\label{CNOT}
%\end{align}
%In matrix representation, 
it is represented by the unitary matrix,
\begin{equation}
U_{\te{CNOT}}=\begin{pmatrix}
1 & 0 & 0 & 0\\
0 & 1 & 0 & 0\\
0 & 0 & 0 & 1\\
0 & 0 & 1 & 0 
\end{pmatrix}.
\label{Uni_CNOT}
\end{equation}
Thus, the CNOT gate operation maps $|\Psi_i\rangle\to|\Psi_{\tau}\rangle = U_{\te{CNOT}}|\Psi_i\rangle$. Notice that $U_{\te{CNOT}}$ does not couple the sectors $\{|00\rangle,|01\rangle\}$ and $\{|10\rangle,|11\rangle\}$, and only in the latter we have the nontrivial operation, $|10\rangle\leftrightarrow|11\rangle$.

 In the present work, we consider a dynamical implementation of the CNOT gate, i.e. $|10\rangle\leftrightarrow|11\rangle$, as described by Eq. (\ref{Uni_CNOT}). To this end, we design a Hamiltonian $H_{\te{CNOT}}(t)$ that contains only locally driven terms, thus enabling control by local external magnetic fields. We require that the gate operation $|10\rangle\leftrightarrow|11\rangle$ is performed adiabatically in the instantaneous ground-state of $H_{\te{CNOT}}(t)$, resembling the methods of adiabatic quantum computation~\cite{RevModPhys.90.015002}. This feature also guarantees the robustness of the implementation while imposing certain additional restrictions. In particular, the adiabaticity condition imposes that the driving time $\tau$ to go, for example, from $|10\rangle$ to $|11\rangle$ is much longer than any other internal timescale of the system \cite{STA_review,duncan2025}. If $\epsilon$ is the smallest energy gap, this means $\tau \gg \epsilon^{-1}$ (throughout this work we consider $\hbar\equiv 1$). 

For finite driving times $\tau\sim \epsilon^{-1}$, the desired operation deteriorates due to transitions to higher energy states in the spectrum. To correct for these undesired excitations, we introduce an additional control field, the counterdiabatic Hamiltonian~\cite{Demirplak2003-ee,Berry_2009,Campo_PRL_2013,STA_review,Nakahara_2022,Ieva_PRX_2023,Polkovnikov_PRB_2024,duncan2025,Tancara_NPJ_2025}, which restores the gate operation for arbitrary $\tau$~\cite{Carolan_2023}. Remarkably, even in the presence of environmental noise, we show that the dynamics generated by the bare Hamiltonian $H_{\text{CNOT}}(t)$ plus the counterdiabatic term accurately describes Eq.~\eqref{Uni_CNOT}, revealing an interesting and system-independent trade-off between noise strength and driving times.

\section{Hamiltonian gates}

In a more physically appealing description, and for closed quantum systems, the output state $|\Psi_{\tau}\rangle$ resulting from the application of a gate $U$ in a given initial state $|\Psi_i\rangle$ can be seen as the result of a time evolution dictated by some controllable time-dependent Hamiltonian $H(t)$ \cite{Hen_2014,Hen_2015,Santos_2018}.

A natural way to build the Hamiltonian $H(t)$ that corresponds to a desired unitary $U$ is by using some inverse engineering method \cite{Berry_2009,PRL_STA_har_2010,SARANDY2011,Kang2016-zn,Santos_2018,STA_review}. For unitary dynamics, we can always write $H(t)=i\dot U(t)U^{\dagger}(t)$, with $U(t)$ satisfying $U(t)U^{\dagger}(t)=\mathbf 1$ and $U(0)=\mathbf 1$ \cite{nielsen2000quantum}. These two conditions can be used to define a physical $U(t)$, from which the associated Hamiltonian can be obtained directly (in fact, the gauge symmetry $|\tilde \Psi(t)\rangle\to V(t)|\Psi(t)\rangle$, for some unitary operator $V(t)$, allows us to build different but equivalent $H(t)$). For example, following the general algorithm introduced in Ref.\cite{Santos_2018}, it is possible to show that
\begin{eqnarray}
    H(t)=-\frac{\dot \varphi(t)}{4}\left(\sigma^z_1-\mathbf 1\right)\left(\sigma^x_2-\mathbf 1\right),
    \label{Alan_CNOT}
\end{eqnarray}
describes the CNOT operation (\ref{Uni_CNOT}) at time $\tau$, that is, $U_{\te{CNOT}}=\mc T_{>}\exp\left(-i\int_0^\tau dt\:H(t)\right)$, where $\mc T_{>}$ is the time ordering operator (which does not play a role here since $[H(t),H(t')]=0$). Moreover, $\varphi(t)$ is an arbitrary function satisfying the boundary conditions $\varphi(0)=2n\pi$ and $\varphi(\tau)=(2n+1)\pi$, with $n$ being an integer, $\sigma^{a=x,y,z}_j$ are the Pauli matrices for the qubit $j$, and $\mathbf 1$ is the $2\times2$ identity matrix. Notice the non-usual, $ZX$, form of the qubit-qubit interaction in Eq.~(\ref{Alan_CNOT}). Such sort of $ZX$ interaction can be implemented using cross-resonance gates with superconducting qubits \cite{Rigetti_PRB_2010,Chow_PRL_2010,Sundaresan_PRX_2020,Dogan_PRA_2023}. Interestingly, this form of interaction also arises in certain two-dimensional Kitaev quantum spin liquid candidates\cite{PRL_gamma_Int_2014}. 

Although the Hamiltonian (\ref{Alan_CNOT}) describes the desired CNOT gate (\ref{Uni_CNOT}) in finite time $\tau$, it requires control in the qubit–qubit $ZX$ interaction, which adds both complexity and resource demands relative to simple local single-qubit controls. In addition to this, by choosing a linear ramp $\varphi(t)=\pi t/\tau$, we obtain a trivial time-independent Hamiltonian. %In fact, there are other systematic approaches to the problem of finding time-independent Hamiltonians for quantum gates utilizing machine learning methods~\cite{Innocenti_2020}. 
In the present work, we follow a different path and devise driven Hamiltonians that implement quantum gates to highlight the importance of external local control for fast gate times. In the next section, we will provide an alternative to Eq.~(\ref{Alan_CNOT}), whose \textit{ground-state dynamics} describes the operation $|10\rangle\leftrightarrow|11\rangle$.

\begin{figure}
    \centering
    \includegraphics[width=1.0\linewidth]{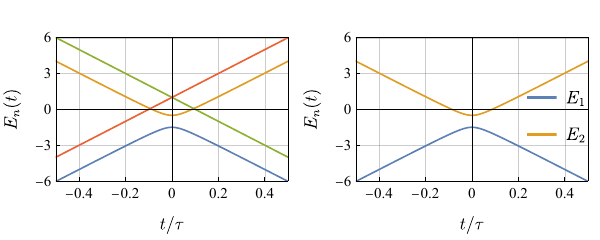}
    \caption{Energy spectrum of the Hamiltonian in Eq. (\ref{H_main}) for a linear drive of the second qubit, $J_2(t)=J_2t/\tau$, with $t\in[-\tau/2,\tau/2]$. Left: full spectrum. Right: spectrum in the nontrivial sector $\{|10\rangle,|11\rangle\}$. The energy gap between $E_2(t)$ and $E_1(t)$ is $\Delta(t)=2\alpha_+(t)$, with a minimum value $\Delta(0)=2g$. Parameters are: $J_2=10J_1$, $g=0.5J_1$, and $J_1=1$.}
    \label{fig_Energies}
\end{figure}

\section{CNOT Hamiltonian}

%We seek a Hamiltonian $H_{\te{CNOT}}(t)$ with the following two properties:
%\begin{itemize}
%    \item $H_{\te{CNOT}}(t)$ contains only control under individual qubits
%    \item The operation $|10\rangle\leftrightarrow|11\rangle$ is performed in the instantaneous ground-state of $H_{\te{CNOT}}(t)$.
%\end{itemize}
%As mentioned previously, the first requirement relies on the experimental feasibility of $H_{\te{CNOT}}(t)$, while the second allows one to realize adiabatic quantum computation. 

\begin{figure*}
     \centering
     \begin{subfigure}
         \centering
         \includegraphics[width=0.317\textwidth]{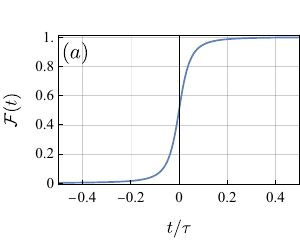}
         %\caption{$y=x$}
         %\label{fig:y equals x}
     \end{subfigure}
     \hfill
     \begin{subfigure}
         \centering
         \includegraphics[width=0.325\textwidth]{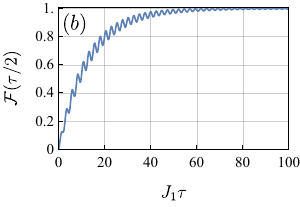}
         %\caption{$y=3\sin x$}
         %\label{fig:three sin x}
     \end{subfigure}
     \hfill
     \begin{subfigure}
         \centering
         \includegraphics[width=0.325\textwidth]{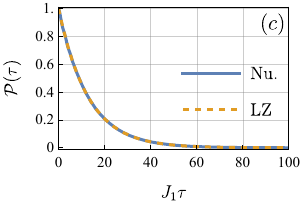}
         %\caption{$y=5/x$}
         %\label{fig:five over x}
     \end{subfigure}
        \caption{(a) Instantaneous fidelity, $\mc F(t)=|\langle\Psi(t)|11\rangle|^2$ with $|\Psi(0)\rangle =|E_1(-\tau/2)\rangle$, between the target state $|11\rangle$ and the evolved state $|\Psi(t)\rangle$ in the adiabatic regime $\tau=200J^{-1}_1$. (b) Final fidelity $\mc F(\tau/2)$ as a function of the driving time $\tau$. (c) Final transition probability $\mc P(\tau)$ as a function of $\tau$. The blue line represents the numerical solution while the dashed orange line the Landau-Zener formula in Eq. (\ref{LZ_formula}). Here we considered $J_2(t)=J_2t/\tau$. The others parameters are: $J_2=10J_1$, $g=0.5J_1$ and $J_1=1$. }
        \label{fig_fidelity}
\end{figure*}

We consider the following two-qubit Hamiltonian,
\begin{eqnarray}
    H_{\te{CNOT}}(t)=J_1\sigma^z_1 + J_2(t)\sigma^z_2 + \frac{g}{2}(\sigma^z_1-\mathbf1)\sigma^x_2.
    \label{H_main}
\end{eqnarray}
Here, $J_1>0$ is an overall energy scale, $J_2(t)$ is a time-dependent magnetic field driving the second qubit and $g>0$ gives the $ZX$ qubit-qubit interaction strength. In the computational basis, the Hamiltonian can be written as
\begin{equation}
H_{\te{CNOT}}(t)=\begin{pmatrix}
K_+(t) & 0 & 0 & 0\\
0 & K_-(t) & 0 & 0\\
0 & 0 & -K_-(t) & -g\\
0 & 0 & -g & -K_+(t) 
\end{pmatrix},
\label{H_CNOT_matrix}
\end{equation}
where $K_{\pm}(t)\equiv J_1\pm J_2(t)$. We can clearly see that, as in the unitary operator (\ref{Uni_CNOT}), there is no coupling between sectors $\{|00\rangle,|01\rangle\}$ and $\{|10\rangle,|11\rangle\}$ in the above Hamiltonian. This implies that, if the initial state $|\Psi_i\rangle$ belongs to one of these sectors, the evolved final state $|\Psi_{\tau}\rangle$ will remain in that sector. As we shall see shortly, this fact allows us to simplify the time evolution.

The instantaneous eigenstates and eigenenergies of $H_{\te{CNOT}}(t)$ can be obtained directly. The instantaneous eigenenergies are,
\begin{eqnarray}
    E_1(t)&=&-\alpha_+(t)-K_-(t),\nonumber\\
    E_2(t)&=&\alpha_+(t)-K_+(t),\nonumber\\
    E_3(t)&=&K_-(t),\:\:\:\:E_4(t)\:\:=\:\:K_+(t),
    \label{ener_spec}
\end{eqnarray}
while the corresponding instantaneous eigenstates,
\begin{eqnarray}
    |E_1(t)\rangle&=&\left(0, 0,-\frac{\alpha_-(t)}{\sqrt{g^2+\alpha^2_-(t)}},\frac{g}{\sqrt{g^2+\alpha^2_-(t)}}\right)^{\text{T}},\nonumber\\
    |E_2(t)\rangle&=&\left(0, 0,-\frac{\alpha_+(t)}{\sqrt{g^2+\alpha^2_+(t)}},\frac{g}{\sqrt{g^2+\alpha^2_+(t)}}\right)^{\text{T}},\nonumber\\
    |E_3(t)\rangle&=&|01\rangle,\:\:\:\:\:\:|E_4(t)\rangle\:\:=\:\:|00\rangle,
    \label{eigenstates}
\end{eqnarray}
where $\alpha_{\pm}(t)\equiv J_2(t)\pm\sqrt{g^2+J^2_2(t)}$. In Fig.~\ref{fig_Energies}, we plot the energy spectrum (\ref{ener_spec}) assuming, without loss of generality, linear driving of the second qubit, $J_2(t)=J_2t/\tau$, with $t\in[-\tau/2,\tau/2]$ (negative times here mean the direction of the applied field $J_2(t)$). The crossings seen in Fig.~\ref{fig_Energies}(a) occur between energy levels from the decoupled sectors. As a result, for the gate operation $|10\rangle\leftrightarrow|11\rangle$, such crossings do not exist, as explicitly shown in Fig.~\ref{fig_Energies}(b).

We observe that $|E_1(t)\rangle$ is the instantaneous ground-state of the system and, in particular, that it has the following properties when $|J_2|\gg J_1,g$:
\begin{eqnarray}
    |E_1(-\tau/2)\rangle&\simeq&|10\rangle,\:\:\:\:|E_1(+\tau/2)\rangle\:\:\:\simeq\:\:\:|11\rangle,
\end{eqnarray}
for $J_2>0$, and
\begin{eqnarray}
    |E_1(-\tau/2)\rangle&\simeq&|11\rangle,\:\:\:\:|E_1(+\tau/2)\rangle\:\:\:\simeq\:\:\:|10\rangle,
\end{eqnarray}
for $J_2<0$. Thus, we conclude that the desired gate operation $|10\rangle\leftrightarrow|11\rangle$ can be performed entirely in the ground-state of $H_{\te{CNOT}}(t)$ as long as we have an adiabatic evolution ($\tau\to\infty$) initiated in $|E_1\rangle$, and $J_2(t)$ changes sign with $|J_2(\pm\tau/2)|\gg J_1,g$.

\section{Solving the dynamics} 

Let us now make the previous statement more precise. Consider that at time $t=-\tau/2$ we have $|\Psi_i\rangle=|E_1(-\tau/2)\rangle\simeq|10\rangle$ ($J_2>0$), and we then start to drive the second qubit according to $J_2(t)=J_2t/\tau$ until $t=\tau/2$. The evolved state at time $t$ is $\rho(t)=|\Psi(t)\rangle\langle\Psi(t)|$, where $|\Psi(t)\rangle$ is obtained by solving the Schrödinger equation subject to this initial condition. 

As a measure of the distance between the quantum state $\rho(t)$ and the desired target state $\rho_d\equiv|11\rangle\langle 11|$, we can consider the Uhlmann fidelity \cite{nielsen2000quantum} $\mc F(t)=\left(\Tr[\sqrt{\sqrt{\rho_d}\rho(t)\sqrt{\rho_d}}]\right)^2$. Given that both quantum states are pure, we simply have $\mc F(t)=|\langle\Psi(t)|11\rangle|^2$.
%Note that $\mc F(t)$ is bounded between $0$ and $1$, with $\mc F=0$ when there is no overlap between the states, i.e., when they are orthogonal, and $\mc F=1$ when the two states are identical. Thus, the operation $|10\rangle\leftrightarrow|11\rangle$ is performed successfully if at the end $\mc F(\tau/2)=1$. 
In Fig. \ref{fig_fidelity}(a) we show $\mc F(t)$ in the adiabatic regime, $\tau\to\infty$, while in Fig. \ref{fig_fidelity}(b) we plot the final fidelity $\mc F(\tau/2)$ as a function of $\tau$. As anticipated, we can see that the CNOT operation is perfectly realized for very slow driving.

We now return to the decoupled sectors mentioned above. As we can see from Eq.~(\ref{eigenstates}), $|\Psi_i\rangle$ belongs to $\{|10\rangle,|11\rangle\}$, and therefore $|\Psi(t)\rangle$ does as well. This means that the dynamics occurs in a restricted, two-dimensional Hilbert space $\{|10\rangle,|11\rangle\}$ and is described by the effective two-level Hamiltonian $H_{\te{eff}}(t)=J_2(t)\sigma^z-g\sigma^x-J_1\mathbf 1$. Here, $\sigma^{z,x}$ stand for the Pauli matrices acting in this subspace. The above Hamiltonian is the Landau-Zener (LZ) Hamiltonian apart from the constant $-J_1\mathbf 1$ \cite{Zener}. Thus, remarkably, the predictions for the Landau-Zener model are also valid for our case. In particular, the transition probability to $|E_2(\tau/2)\rangle$, that is, $\mc P(\tau)\equiv|\langle\Psi(\tau/2)|E_2(\tau/2)\rangle|^2$, when $|J_2|\gg J_1,g$, should be given by the LZ formula \cite{Zener,Soriani,Glasbrenner_2023},
\begin{eqnarray}
    \mc P_{\te{LZ}}(\tau)=\exp\left(-\frac{\pi g^2\tau}{J_2}\right).
    \label{LZ_formula}
\end{eqnarray}
In Fig. \ref{fig_fidelity}(c), we compare the analytical expression above with numerical results and observe perfect agreement between them. We believe that this showcases an important advantage of the approach presented in this work, since it allows us to treat a two-qubit problem with an effective \textit{single-qubit} Hamiltonian. This reduces the complexity of the quantum process in experimentally relevant scenarios.

%So far we have seen that the proposed Hamiltonian in Eq. (\ref{H_main}) describes the CNOT operation when $\tau\to\infty$ and $|J_2|\gg J_1,g$. 

In the next section, we will describe and \textit{correct} potential sources of errors in the dynamical implementation of the gate operation with Eq.~(\ref{H_main}). 

\section{Dynamical error mitigation}

To perform the CNOT operation through the dynamics generated by $H_{\te{CNOT}}(t)$ in Eq. (\ref{H_main}), the state $|E_{1}(t)\rangle$ must be driven as perfectly as possible from $|10\rangle$ to $|11\rangle$. To achieve this, we must guarantee an extremely slowly varying $J_{2}(t)$ and the boundary conditions $|J_{2}(\pm\tau/2)|\gg J_{1},\,g$. In the following, we analyze these requirements by considering a finite-time driving in the presence of external noise.
%
%Two of the main assumptions for the Hamiltonian $H_{\te{CNOT}}(t)$ in Eq. (\ref{H_main}) describe the CNOT operation $|10\rangle\leftrightarrow|11\rangle$ are that we perform an infinite long adiabatic evolution $\tau\to\infty$, which guarantees that we remain in the instantaneous ground-state $|E_1(t)\rangle$, and that the applied magnetic field $J_2(t)$ perfectly drives the second qubit under the condition $|J_2|\gg J_1,g$. In the following, we confront these assumptions by considering a finite-time evolution and that control over $J_2(t)$ introduce noise on the system. 

\begin{figure}
    \centering
    \includegraphics[width=0.9\linewidth]{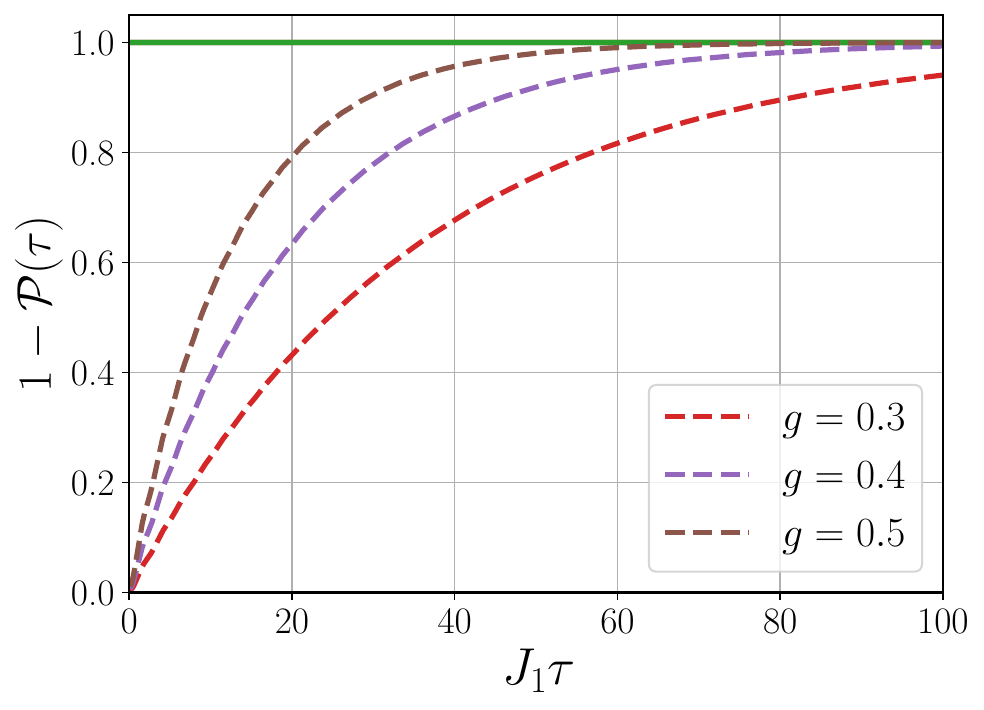}
    \caption{Probability to end up in the ground-state $|E_1(\tau/2)\rangle$ with (continuous lines) and without (dashed lines) the counterdiabatic Hamiltonian in Eq. (\ref{Hcd_cnot}). Results for $g=0.3,0.4,0.5$ (in units of $J_1$) and $J_2(t)=J_2t/\tau$ are shown. In the case with $H_{\te{CD}}(t)$ all three curves for different $g$ collapse on top of each other. Parameters are: $J_2=10J_1$ and $J_1=1$.}
    \label{fig_fidelity_cd}
\end{figure}

\subsection{Finite-time control} 

From the results shown in Fig. \ref{fig_fidelity}, it is clear that the implementation of the CNOT gate by means of Eq. (\ref{H_main}) is highly degraded for small $\tau$, i.e., far from the adiabatic regime. More precisely, the condition for an adiabatic evolution is $\tau\gg g^{-1}$, noting that $g$ sets the energy gap between $E_2(t)$ and $E_1(t)$ (see Eq. (\ref{ener_spec}) and Fig. \ref{fig_Energies}). The physical meaning of this condition is straightforward: a finite-time evolution only induces transitions between eigenstates within an energy width $\tau^{-1}$. When $g\le\tau^{-1}$, such transitions are likely. Far from this limit, that is, $\tau\gg g^{-1}$, excitations are rather suppressed and adiabaticity is ensured.

The adiabatic limit discussed above can be achieved exactly for \textit{arbitrarily small} $\tau$ considering the application of an additional control field $H_{\te{CD}}(t)$ \cite{Berry_2009,STA_review},
\begin{eqnarray}
    H_{\te{CD}}(t)=i\sum_{m\neq n}\frac{\langle E_m(t)|\dot H(t)|E_n(t)\rangle}{E_n(t)-E_m(t)}|E_m(t)\rangle\langle E_n(t)|,
    \label{Berry_Hcd}
\end{eqnarray}
where $\dot H(t)=\frac{d}{dt}H(t)$ is the reference dynamics (in our case $H(t)\equiv H_{\te{CNOT}}(t)$) and $|E_n(t)\rangle$ and $E_n(t)$ are the instantaneous eigenstates and eigenenergies of $H(t)$, respectively. The dynamics generated by $H(t)+H_{\te{CD}}(t)$ remains adiabatic for any $\tau$ as long as the system spectrum is gapped during the time evolution, $\Delta(t)\neq 0$. Although it seems that we obtain a general \textit{adiabatic} speed-up for free, this process becomes highly costly as $\tau\to 0$ \cite{Santos_2015,Steve_2017,Hu_2018,Puebla2020PRR,Carolan_2023}. Precisely, it was shown in \cite{Santos_2015} that for a two-qubit problem the cost scales as $\tau^{-1}$ when $\tau\to 0$.

Using the results in Eqs. (\ref{ener_spec}) and (\ref{eigenstates}), the counterdiabatic Hamiltonian in Eq. (\ref{Berry_Hcd}) for our case is simply given by, 
\begin{eqnarray}
    H_{\te{CD}}(t)=-\frac{g\dot J_2(t)}{4\left[g^2+J^2_2(t)\right]}(\sigma^z_1-\mathbf1)\sigma^y_2.
    \label{Hcd_cnot}
\end{eqnarray}
The time-dependent prefactor in the above equation is exactly the same as that appearing in the corresponding $H_{\te{CD}}(t)$ for the LZ model \cite{Carolan_PRA_2020}. Note also that $H_{\te{CD}}(t)$ is entirely spanned in the sector $\{|10\rangle,|11\rangle\}$. This, once more, highlights the fact that we have an effective LZ problem. In Refs. \cite{Zhang_PRL_2013,Hu_2018,Zhang_PRL_2024}, the counterdiabatic dynamics of the LZ model was experimentally realized.

In Fig.~\ref{fig_fidelity_cd}, we show $1-\mc P(\tau)$, that is, the probability to end up in the ground-state $|E_1(\tau/2)\rangle$ for the cases with and without the counterdiabatic Hamiltonian (\ref{Hcd_cnot}). Indeed, we observe that the introduction of $H_{\te{CD}}(t)$ guarantees the adiabaticity of the evolution and thus makes the gate operation successful for arbitrarily small $\tau$, as can be seen by the solid line in the graph. The dashed lines in Fig.~\ref{fig_fidelity_cd}, displays our results for different values of $g$ in the absence of counterdiabatic control. In such cases, the larger $g$, the faster we get to $\mc P(\tau) \to 0$ for a given $\tau$. However, it is important to remember that by increasing $g$ and keeping $J_2(\pm\tau/2)$ fixed, we increase the distance between $|E_{1}(\tau/2)\rangle$ and the state $|11\rangle$, which spoils the operation of the gate. On the other hand, in the presence of $H_{\te{CD}}(t)$, we can see that the gap does not play a substantial role, as long as it is nonzero.  

\subsection{Noise field}

As a second possible source of errors, we consider the effects of a noise that acts on our system through fluctuations in the external driving field. It is natural and common to expect the presence of such stochastic noises with certain statistical properties on experimental platforms~\cite{Kiely_2021}. Mathematically, it is modeled by replacing $J_2(t)$ with $J_2(t)+\eta(t)$ in Eq.~(\ref{H_main}), where $\eta(t)$ describes the noise (with zero mean) and from now on $J_{2}(t)$ must be understood as the averaged controllable part of the external driving. For this reason, the counterdiabatic Hamiltonian (\ref{Hcd_cnot}) maintains its form. This more realistic two-qubit CNOT Hamiltonian then reads,
\begin{eqnarray}
    \mc H(t)=H_{\te{CNOT}}(t)+\eta(t)\sigma^z_2.
\end{eqnarray}
The meaningful quantity now that describes the state of the system at a given time $t$ is the noise-averaged density operator $\rho(t)\equiv \overline{\rho_{\eta}(t)}$, where $\rho_{\eta}(t)$ is the density operator for a given realization of the noise and $\overline{[\bullet]}$ means the ensemble average. For Gaussian white noise, the density operator $\rho(t)$ is known to satisfy the following Markovian master equation \cite{Kiely_2021},  
\begin{eqnarray}
    \dot\rho(t)=-i[H_{\te{CNOT}}(t),\rho(t)]+\alpha\left[\sigma^z_2\rho(t)\sigma^z_2-\rho(t)\right].
    \label{ME}
\end{eqnarray}
The second term in the above equation, that is, the dissipative term, encodes the stochasticity of the noise process, and $\alpha>0$ measures its strength. Notice that this equation is in the Lindblad form, with only one quantum jump operator given by $L=\sqrt{\alpha}\sigma^z_2$. Additionally, this same equation also describes the continuous monitoring of $\sigma^z_2$ at rate $\alpha$ \cite{breuer2002theory}. In Ref. \cite{Hu_2018}, by using a \ce{^{171}Yb+} trapped ion as a qubit, the authors experimentally realize the dissipative dynamics of LZ model described by the above master equation.

\begin{figure}
     \centering
     \begin{subfigure}
         \centering
         \includegraphics[width=7cm, height=7.0cm]{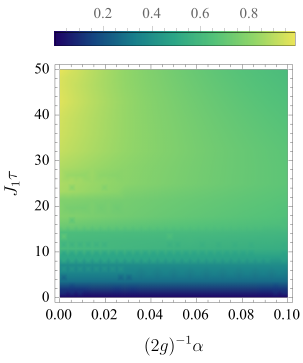}
         %\caption{$y=x$}
         %\label{fig:y equals x}
     \end{subfigure}
     \hfill
     \begin{subfigure}
         \centering
         \includegraphics[width=0.38\textwidth]{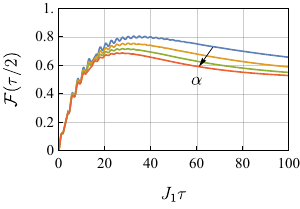}
         %\caption{$y=3\sin x$}
         %\label{fig:three sin x}
     \end{subfigure}
        \caption{Fidelity $\mc F(\tau/2)$ between the target state $|11\rangle$ and the evolved state $\rho(\tau/2)$ in the presence of a weak Gaussian white noise of strength $\alpha$. Upper panel: density plot of $\mc F(\tau/2)$ in the $(\alpha,\tau)$ plane. Lower panel: $\mc F(\tau/2)$ as a function of $\tau$ for $\alpha=0.04,0.06,0.08,0.1$ (in units of the gap $\Delta(0)=2g$). Here we considered $J_2(t)=J_2t/\tau$. Parameters are: $J_2=10J_1$, $g=0.5J_1$, and $J_1=1$. }
        \label{fig_fidelity_noise1}
\end{figure}

As in the noise-free case, the figure of merit to consider is again the fidelity $\mc F(t)$ between $\rho(t)$ and $\rho_d=|11\rangle\langle 11|$, with $\rho(0)=|10\rangle\langle10|$. However, given the mixedness of $\rho(t)$ in the presence of the noise, we now have $\mc F(t)=\Tr[\rho_d\rho(t)]=\langle11|\rho(t)|11\rangle$. Figure~\ref{fig_fidelity_noise1} presents our results for the fidelity of the gate operation at the end of the time evolution, i.e. $t=\tau/2$, and in the weak noise regime.
Focusing on the upper panel, we observe that the noise strongly affects the performance of the gate operation, even for $\tau\to\infty$, spoiling the results we have obtained previously with unitary evolution [see Fig. \ref{fig_fidelity}(b)]. For example, for $\alpha=0.08g$, the highest fidelity is around $0.8$, as can be seen in the lower panel of Fig.~\ref{fig_fidelity_noise1}. The only situation where we have $\mc F(\tau/2)=1$ is strictly when $\alpha=0$ and $\tau\to\infty$. 

The lower values are also accompanied by a nonmonotonic behavior of $\mc F(\tau/2)$ as a function of $\tau$, which reveals an optimal driving time, $\tau^*$, where $\mc F(\tau/2)$ is maximum. This optimal time decreases as $\alpha$ increases. As $\tau$ continues to increase, we see that $\mc F(\tau/2)$ approaches a constant value. This can be understood in terms of the fixed point of Eq. (\ref{ME}): its steady state, $\dot\rho(t)=0$, is the infinite temperature state $\rho(\infty)=\mathbf 1/2$, which makes $\mc F(\infty)=\langle11|\rho(\infty)|11\rangle=1/2$. Thus, a long-time evolution does not imply a better performance in this more realistic scenario with noise.

\begin{figure}
     \centering
     \begin{subfigure}
         \centering
         \includegraphics[width=7cm, height=7.0cm]{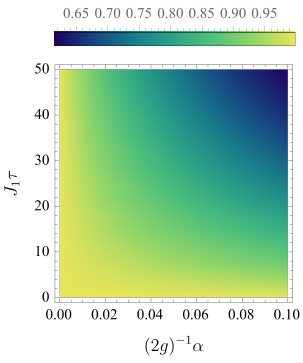}
         %\caption{$y=x$}
         %\label{fig:y equals x}
     \end{subfigure}
     \hfill
     \begin{subfigure}
         \centering
         \includegraphics[width=0.38\textwidth]{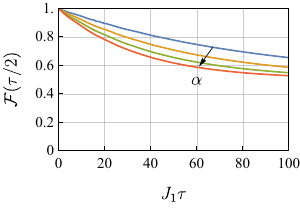}
         %\caption{$y=3\sin x$}
         %\label{fig:three sin x}
     \end{subfigure}
        \caption{Fidelity $\mc F(\tau/2)$ between the target state $|11\rangle$ and the evolved state $\rho(\tau/2)$ in the presence of a weak Gaussian white noise of strength $\alpha$ and considering the counterdiabatic field in Eq. (\ref{Hcd_cnot}). Upper panel: density plot of $\mc F(\tau/2)$ in the $(\alpha,\tau)$ plane. Lower panel: $\mc F(\tau/2)$ as a function of $\tau$ for $\alpha=0.04,0.06,0.08,0.1$ (in units of the gap $\Delta(0)=2g$). Here we considered $J_2(t)=J_2t/\tau$. Parameters are: $J_2=10J_1$, $g=0.5J_1$ and $J_1=1$. }
        \label{fig_fidelity_noise2}
\end{figure}

From the lower panel of Fig. \ref{fig_fidelity_noise1} we conclude that, for the set of parameters chosen, the noise effects start to dominate the unitary evolution around the optimal time $\tau^* \approx30J^{-1}_1$. When evaluated together with the fundamental barrier of $\mc F(\tau\to\infty)=1/2$, this tells us that the only way to achieve better performance is by keeping $\tau$ finite and relatively small, i.e., $\tau<30J^{-1}_1$. Knowing the fact that the counterdiabatic Hamiltonian (\ref{Hcd_cnot}) strongly suppresses excitations in short times, $H_{\te{CD}}(t)\sim \tau^{-1}$, we will now change the unitary part of (\ref{ME}) to $H_{\te{CNOT}}(t)+H_{\te{CD}}(t)$. Although this ignores a possible (and probable) noise dependence that comes from the control of $H_{\te{CD}}(t)$, the fact that $H_{\te{CD}}(t)\sim \tau^{-1}$ guarantees that the additional noise dependence would be rather suppressed by the counterdiabatic unitary evolution in short times and weak noise limit.  

In Fig.~\ref{fig_fidelity_noise2}, we show the results for the case in which the counterdiabatic Hamiltonian is added. As we can clearly see in both panels, the nonunitary dynamics with $H_{\te{CD}}(t)$  produces excellent results for $\mc F(\tau/2)$ in short times $\tau<30J^{-1}_1$ even for a reasonable noise strength of $\alpha= 0.2g$. Therefore, we conclude that the counterdiabatic Hamiltonian (\ref{Hcd_cnot}) can avoid noisy errors in the gate implementation, especially for very short time evolutions.

It is worth mentioning the trade-off between $\tau$ and $\alpha$ seen in Fig.~\ref{fig_fidelity_noise2}. We see that, to achieve high fidelity, long-time evolutions require very weak noise, whereas strong noise requires short-time evolutions. This implies that, only in the region
\begin{eqnarray}
    \tau\alpha&\lesssim&1,
    \label{gate_inequ}
\end{eqnarray}
we will have $\mc F(\tau/2)\approx 1$ and consequently the desired gate operation will be performed close to perfect. The above relation can be understood again due to the fact that $H_{\te{CD}}(t)\sim\tau^{-1}$. When $H_{\te{CD}}(t)$ is added, the unitary part of Eq. (\ref{ME}) scales as $\tau^{-1}$ [$H_{\te{CNOT}}(t)$ does not scale with $\tau$], while the nonunitary part as $\alpha$. Thus, for larger $\tau$ we need to decrease $\alpha$ in order to counteract the effects of noise, and vice versa. Notice that the relationship $H_{\te{CD}}(t)\sim \dot H(t)\sim \tau^{-1}$ is general, i.e., system-independent. Therefore, we expect Eq. (\ref{gate_inequ}) to be valid for other Hamiltonian gates described by a master equation of the form (\ref{ME}) in the presence of the corresponding $H_{\te{CD}}(t)$.

\section{$N$-qubits generalization}
To conclude our analysis, we highlight the fact that the results shown so far can be directly generalized to a $N$-qubit gates. For example, the operation $|11\cdots10\rangle\leftrightarrow|11\cdots11\rangle$, which for $N=3$ corresponds to a Toffoli gate \cite{nielsen2000quantum}, can be adiabatically implemented in the instantaneous ground-state of the following Hamiltonian,
\begin{equation}
    H^{(N)}(t)=J\sum_{i=1}^{N-1}\sigma^z_i+J_N(t)\sigma^z_N-g\left[\prod_{i=1}^{N-1}\frac{\mathbf 1-\sigma^z_i}{2}\right]\sigma^x_N,
    \label{NHamil}
\end{equation}
as long as $|J_N(\pm\tau/2)|\gg J,g$. The counterdiabatic control field (\ref{Berry_Hcd}) corresponding to the Hamiltonian above can be determined as,
\begin{equation}
    H^{(N)}_{\te{CD}}(t)=\frac{g\dot J_{N}(t)}{2[g^2+J^2_N(t)]}\left[\prod_{i=1}^{N-1}\frac{\mathbf 1-\sigma^z_i}{2}\right]\sigma^y_N.
    \label{NHcd}
\end{equation}
Once again, we have an effective LZ problem. This represents a significant advantage in designing the qubit–qubit couplings necessary to implement the gate operation $|11\cdots10\rangle\leftrightarrow|11\cdots11\rangle$ in finite time \cite{Santos_2015}.

\section{Concluding remarks} In this work, we have considered the problem of dynamically implementing the CNOT operation $|10\rangle\leftrightarrow|11\rangle$. By locally driving a system of two-interacting qubits, we have shown that the desired gate operation can be performed adiabatically in terms of an effective Landau-Zener (two-level) problem. More realistic situations where errors arise have been analyzed, such as when the gate operation is subjected to non-adiabatic and noise effects. We have shown that counterdiabatic control can properly avoid these sources of errors and thus that the control theory based on Eq.~(\ref{Hcd_cnot}) is relatively robust against weak noise for short times $\tau$.

Finding new ways to protect unitary gate operations against different forms of errors is essential to quantum computing. Here, we have provided a simple and general way to do this on the basis of shortcuts to adiabaticity. In spirit, our approach is similar to what was proposed in Ref. \cite{Santos_2015}. However, in our analysis, we have taken an additional step forward and also shown the effectiveness of counterdiabatic control in mitigating errors introduced by possible classical noise in the control fields and the presence of environmental effects. As a future work, it would be interesting to make a comprehensive comparative analysis of the performance of different quantum control methods~\cite{Steve2025} in achieving noise robust quantum gates.

\acknowledgments{M.F.C would like to thank E. Doucet for insightful discussions on the theory of open quantum systems, and R. Robertson for drawing our attention to general $N$-qubit gates. This work was supported by Conselho Nacional de Desenvolvimento Cient\'{i}fico e Tecnol\'{o}gico (CNPq), Brazil, through grant No. 200267/2023-0. M.V.S.B. acknowledges the support of CNPq, under Grant No. 304120/2022-7, and the S\~ao Paulo Research Foundation (FAPESP), Brazil, Process Number 2022/15453-0.
S.D. and M.F.C. acknowledge support from the John Templeton Foundation under Grant No. 62422.}

\bibliographystyle{eplbib}
\bibliography{epl_Refs}

\begin{thebibliography}{10}
\expandafter\ifx\csname url\endcsname\relax\def\url#1{\texttt{#1}}\fi

\bibitem{nielsen2000quantum}
\Name{Nielsen M. \and Chuang I.} \Book{Quantum Computation and Quantum Information} Cambridge Series on Information and the Natural Sciences (Cambridge University Press) 2000.

\bibitem{Sanders2017}
\Name{Sanders B.~C.} \Book{How to Build a Quantum Computer} 2399-2891 (IOP Publishing) 2017.

\bibitem{BruzewiczAPR2019}
\Name{Bruzewicz C.~D., Chiaverini J., McConnell R. \and Sage J.~M.} \REVIEW{Appl. Phys. Rev.}{6}{2019}{021314}.

\bibitem{Slussarenko2019APR}
\Name{Slussarenko S. \and Pryde G.~J.} \REVIEW{Appl. Phys. Rev.}{6}{2019}{041303}.

\bibitem{Wintersperger2023}
\Name{Wintersperger K., Dommert F., Ehmer T., Hoursanov A., Klepsch J., Mauerer W., Reuber G., Strohm T., Yin M. \and Luber S.} \REVIEW{EPJ Quantum Technology}{10}{2023}{32}.

\bibitem{Ezratty2023EPJA}
\Name{Ezratty O.} \REVIEW{Eur. Phys. J. A}{59}{2023}{94}.

\bibitem{Tripathi2025CR}
\Name{Tripathi V., Kowsari D., Saurav K., Zhang H., Levenson-Falk E.~M. \and Lidar D.~A.} \REVIEW{Chemical Reviews}{125}{2025}{5745}.

\bibitem{PhysRevA.65.042301}
\Name{Ahn C., Doherty A.~C. \and Landahl A.~J.} \REVIEW{Phys. Rev. A}{65}{2002}{042301}.

\bibitem{PhysRevA.86.012317}
\Name{Gorman D.~J., Young K.~C. \and Whaley K.~B.} \REVIEW{Phys. Rev. A}{86}{2012}{012317}.

\bibitem{Goerz_2014}
\Name{Goerz M.~H., Reich D.~M. \and Koch C.~P.} \REVIEW{New Journal of Physics}{16}{2014}{055012}.

\bibitem{Deffner_2014}
\Name{Deffner S.} \REVIEW{Journal of Physics B: Atomic, Molecular and Optical Physics}{47}{2014}{145502}.

\bibitem{Du_2016}
\Name{Du Y.-X., Liang Z.-T., Li Y.-C., Yue X.-X., Lv Q.-X., Huang W., Chen X., Yan H. \and Zhu S.-L.} \REVIEW{Nature Communications}{7}{2016}{12479}.

\bibitem{Santos_2017}
\Name{Santos A.~C. \and Sarandy M.~S.} \REVIEW{Journal of Physics A: Mathematical and Theoretical}{51}{2017}{025301}.

\bibitem{Utkan2022PRR}
\Name{G\"ung\"ord\"u U. \and Kestner J.~P.} \REVIEW{Phys. Rev. Res.}{4}{2022}{023155}.

\bibitem{Kanaar2024QST}
\Name{Kanaar D.~W. \and Kestner J.~P.} \REVIEW{Quantum Science and Technology}{9}{2024}{035011}.

\bibitem{PRL_URC_2024}
\Name{Poggi P.~M., De~Chiara G., Campbell S. \and Kiely A.} \REVIEW{Phys. Rev. Lett.}{132}{2024}{193801}.

\bibitem{Li_2024}
\Name{Li B., Calarco T. \and Motzoi F.} \REVIEW{npj Quantum Information}{10}{2024}{66}.

\bibitem{Vetter2024}
\Name{Vetter P.~J., Reisser T., Hirsch M.~G., Calarco T., Motzoi F., Jelezko F. \and M{\"u}ller M.~M.} \REVIEW{npj Quantum Information}{10}{2024}{96}.

\bibitem{singh2025}
\Name{Singh J., Reuter J.~A., Calarco T., Motzoi F. \and Zeier R.} \REVIEW{arXiv preprint arXiv:2503.06768}{}{2025}{}.

\bibitem{Acconcia2015PRE}
\Name{Acconcia T.~V., Bonan\ifmmode~\mbox{\c{c}}\else \c{c}\fi{}a M. V.~S. \and Deffner S.} \REVIEW{Phys. Rev. E}{92}{2015}{042148}.

\bibitem{Walelign2024PRA}
\Name{Walelign H.~Y., Cai X., Li B., Barnes E. \and Nichol J.~M.} \REVIEW{Phys. Rev. Appl.}{22}{2024}{064029}.

\bibitem{Amer2025}
\Name{Amer H., Piliouras E., Barnes E. \and Economou S.~E.} \REVIEW{arXiv preprint arXiv:2504.09767}{}{2025}{}.

\bibitem{Hegade_PRA_2021}
\Name{Hegade N.~N., Paul K., Ding Y., Sanz M., Albarr\'an-Arriagada F., Solano E. \and Chen X.} \REVIEW{Phys. Rev. Appl.}{15}{2021}{024038}.

\bibitem{Hegade2022PRR}
\Name{Hegade N.~N., Chen X. \and Solano E.} \REVIEW{Phys. Rev. Res.}{4}{2022}{L042030}.

\bibitem{Zhao2024PRL}
\Name{Zhao H., Bukov M., Heyl M. \and Moessner R.} \REVIEW{Phys. Rev. Lett.}{133}{2024}{010603}.

\bibitem{PRA_UQC_1995}
\Name{Barenco A., Bennett C.~H., Cleve R., DiVincenzo D.~P., Margolus N., Shor P., Sleator T., Smolin J.~A. \and Weinfurter H.} \REVIEW{Phys. Rev. A}{52}{1995}{3457}.

\bibitem{RevModPhys.90.015002}
\Name{Albash T. \and Lidar D.~A.} \REVIEW{Rev. Mod. Phys.}{90}{2018}{015002}.

\bibitem{STA_review}
\Name{Gu\'ery-Odelin D., Ruschhaupt A., Kiely A., Torrontegui E., Mart\'{\i}nez-Garaot S. \and Muga J.~G.} \REVIEW{Rev. Mod. Phys.}{91}{2019}{045001}.

\bibitem{duncan2025}
\Name{Duncan C.~W., Poggi P.~M., Bukov M., Zinner N.~T. \and Campbell S.} \REVIEW{arXiv preprint arXiv:2501.16436}{}{2025}{}.

\bibitem{Demirplak2003-ee}
\Name{Demirplak M. \and Rice S.~A.} \REVIEW{J. Phys. Chem. A}{107}{2003}{9937}.

\bibitem{Berry_2009}
\Name{Berry M.~V.} \REVIEW{Journal of Physics A: Mathematical and Theoretical}{42}{2009}{365303}.

\bibitem{Campo_PRL_2013}
\Name{del Campo A.} \REVIEW{Phys. Rev. Lett.}{111}{2013}{100502}.

\bibitem{Nakahara_2022}
\Name{Nakahara M.} \REVIEW{Philosophical Transactions of the Royal Society A: Mathematical, Physical and Engineering Sciences}{380}{2022}{}.

\bibitem{Ieva_PRX_2023}
\Name{\ifmmode \check{C}\else \v{C}\fi{}epait\ifmmode~\dot{e}\else \.{e}\fi{} I., Polkovnikov A., Daley A.~J. \and Duncan C.~W.} \REVIEW{PRX Quantum}{4}{2023}{010312}.

\bibitem{Polkovnikov_PRB_2024}
\Name{Morawetz S. \and Polkovnikov A.} \REVIEW{Phys. Rev. B}{110}{2024}{024304}.

\bibitem{Tancara_NPJ_2025}
\Name{Tancara D. \and Albarrán-Arriagada F.} \REVIEW{npj Quantum Information}{11}{2025}{}.

\bibitem{Carolan_2023}
\Name{Carolan E., \c{C}akmak B. \and Campbell S.} \REVIEW{Phys. Rev. A}{108}{2023}{022423}.

\bibitem{Hen_2014}
\Name{Hen I.} \REVIEW{Frontiers Phys.}{2}{2014}{}.

\bibitem{Hen_2015}
\Name{Hen I.} \REVIEW{Phys. Rev. A}{91}{2015}{022309}.

\bibitem{Santos_2018}
\Name{Santos A.~C.} \REVIEW{Journal of Physics B: Atomic, Molecular and Optical Physics}{51}{2017}{015501}.

\bibitem{PRL_STA_har_2010}
\Name{Chen X., Ruschhaupt A., Schmidt S., del Campo A., Gu\'ery-Odelin D. \and Muga J.~G.} \REVIEW{Phys. Rev. Lett.}{104}{2010}{063002}.

\bibitem{SARANDY2011}
\Name{Sarandy M., Duzzioni E. \and Serra R.} \REVIEW{Physics Letters A}{375}{2011}{3343}.

\bibitem{Kang2016-zn}
\Name{Kang Y.-H., Chen Y.-H., Wu Q.-C., Huang B.-H., Xia Y. \and Song J.} \REVIEW{Scientific Reports}{6}{2016}{30151}.

\bibitem{Rigetti_PRB_2010}
\Name{Rigetti C. \and Devoret M.} \REVIEW{Phys. Rev. B}{81}{2010}{134507}.

\bibitem{Chow_PRL_2010}
\Name{Chow J.~M., C\'orcoles A.~D., Gambetta J.~M., Rigetti C., Johnson B.~R., Smolin J.~A., Rozen J.~R., Keefe G.~A., Rothwell M.~B., Ketchen M.~B. \and Steffen M.} \REVIEW{Phys. Rev. Lett.}{107}{2011}{080502}.

\bibitem{Sundaresan_PRX_2020}
\Name{Sundaresan N., Lauer I., Pritchett E., Magesan E., Jurcevic P. \and Gambetta J.~M.} \REVIEW{PRX Quantum}{1}{2020}{020318}.

\bibitem{Dogan_PRA_2023}
\Name{Dogan E., Rosenstock D., Le~Guevel L., Xiong H., Mencia R.~A., Somoroff A., Nesterov K.~N., Vavilov M.~G., Manucharyan V.~E. \and Wang C.} \REVIEW{Phys. Rev. Appl.}{20}{2023}{024011}.

\bibitem{PRL_gamma_Int_2014}
\Name{Rau J.~G., Lee E. K.-H. \and Kee H.-Y.} \REVIEW{Phys. Rev. Lett.}{112}{2014}{077204}.

\bibitem{Zener}
\Name{Zener C.} \REVIEW{Proceedings of the Royal Society of London. Series A, Containing Papers of a Mathematical and Physical Character}{137}{1932}{696}.

\bibitem{Soriani}
\Name{Soriani A., Naz\'e P., Bonan\ifmmode~\mbox{\c{c}}\else \c{c}\fi{}a M. V.~S., Gardas B. \and Deffner S.} \REVIEW{Phys. Rev. A}{105}{2022}{042423}.

\bibitem{Glasbrenner_2023}
\Name{Glasbrenner E.~P. \and Schleich W.~P.} \REVIEW{Journal of Physics B: Atomic, Molecular and Optical Physics}{56}{2023}{104001}.

\bibitem{Santos_2015}
\Name{Santos A.~C. \and Sarandy M.~S.} \REVIEW{Scientific Reports}{5}{2015}{15775}.

\bibitem{Steve_2017}
\Name{Campbell S. \and Deffner S.} \REVIEW{Phys. Rev. Lett.}{118}{2017}{100601}.

\bibitem{Hu_2018}
\Name{Hu C.-K., Cui J.-M., Santos A.~C., Huang Y.-F., Sarandy M.~S., Li C.-F. \and Guo G.-C.} \REVIEW{Optics Letters}{43}{2018}{3136}.

\bibitem{Puebla2020PRR}
\Name{Puebla R., Deffner S. \and Campbell S.} \REVIEW{Phys. Rev. Res.}{2}{2020}{032020}.

\bibitem{Carolan_PRA_2020}
\Name{Carolan E., Kiely A. \and Campbell S.} \REVIEW{Phys. Rev. A}{105}{2022}{012605}.

\bibitem{Zhang_PRL_2013}
\Name{Zhang J., Shim J.~H., Niemeyer I., Taniguchi T., Teraji T., Abe H., Onoda S., Yamamoto T., Ohshima T., Isoya J. \and Suter D.} \REVIEW{Phys. Rev. Lett.}{110}{2013}{240501}.

\bibitem{Zhang_PRL_2024}
\Name{Zhang J.-W., Bu J.-T., Li J.~C., Meng W., Ding W.-Q., Wang B., Yuan W.-F., Du H.-J., Ding G.-Y., Chen W.-J., Chen L., Zhou F., Xu Z. \and Feng M.} \REVIEW{Phys. Rev. Lett.}{132}{2024}{213602}.

\bibitem{Kiely_2021}
\Name{Kiely A.} \REVIEW{Europhysics Letters}{134}{2021}{10001}.

\bibitem{breuer2002theory}
\Name{Breuer H. \and Petruccione F.} \Book{The Theory of Open Quantum Systems} (Oxford University Press) 2002.

\bibitem{Steve2025}
\Name{Duncan C.~W., Poggi P.~M., Bukov M., Zinner N.~T. \and Campbell S.} \REVIEW{PRX Quantum}{6}{2025}{040201}.

\end{thebibliography}
\end{document}